# Quantum states with Space-like energy-momentum

by

Dan Solomon

Rauland-Borg
3450 W. Oakton
Skokie, IL

March 22, 2006

**Abstract**

A common assumption in quantum field theory is that the energy-momentum 4-vector of any quantum state must be time-like. It will be proven that this is not the case for a Dirac-Maxwell field. In this case quantum states can be shown to exist whose energy-momentum is space-like.

PACS No. 11.10.-z, 03.70.+k, 03.65.-w

Keywords: negative energy, space-like, energy-momentum.

## 1. Introduction

A common assumption in quantum field theory is that the energy-momentum 4-vector of quantum states must be time-like. (See, for example, page 58 of Haag[1], Section 8.1 of Nishijima[2], and Section 10.7 of Weinberg[3]). While this is certainly the case for non-interacting (free) fields we will show that this is not necessarily the case for interacting fields. In this discussion we will examine a Dirac-Maxwell field. For this case it will be shown that there exist quantum states with space-like energy momentum.

Let $\hat{H}$ be the Hamiltonian operator and $\hat{\vec{P}}$ the momentum operator. The energy expectation value of a normalized state vector $|\Omega\rangle$ is $\langle\Omega|\hat{H}|\Omega\rangle$ and the momentum expectation value is $\langle\Omega|\hat{\vec{P}}|\Omega\rangle$. The energy-momentum of the state $|\Omega\rangle$ is then,

$$G(|\Omega\rangle) \equiv \langle\Omega|\hat{H}|\Omega\rangle^2 - \left|\langle\Omega|\hat{\vec{P}}|\Omega\rangle\right|^2 \qquad (1)$$

If $G \geq 0$ the energy-momentum is said to be time-like. If $G < 0$ the energy-momentum is space-like. As stated above, G is generally assumed to greater than or equal to zero (i.e. time-like) for all possible states. It is this assumption that will be challenged in this paper.

Define the operator,

$$\hat{F} = \hat{H} - \hat{\vec{P}} \cdot \vec{e}_1 \qquad (2)$$

where $\vec{e}_1$ is a unit vector in the $x_1$-direction. Let $|\Omega\rangle$ be a normalized state vector. If the energy-momentum of all states is time-like then the following condition must hold,

$$\langle\Omega|\hat{F}|\Omega\rangle \geq 0 \;\forall\; |\Omega\rangle \qquad (3)$$

To show that the above relationship is necessary for the energy-momentum to be time-like consider what would happen if there existed a state $|\Omega'\rangle$ for which the above relationship did not hold, i.e. $\langle\Omega'|\hat{F}|\Omega'\rangle$ was negative. In this case we obtain,

$$\langle\Omega'|\hat{H}|\Omega'\rangle < \langle\Omega'|\hat{\vec{P}}\cdot\vec{e}_1|\Omega'\rangle \tag{4}$$

Since we assume that the energy of any state is positive with respect to the vacuum state, so that $\langle\Omega'|\hat{H}|\Omega'\rangle \geq 0$, we can use the above to obtain,

$$0 > \langle\Omega'|\hat{H}|\Omega'\rangle^2 - \langle\Omega'|\hat{\vec{P}}\cdot\vec{e}_1|\Omega'\rangle^2 \geq G\left(|\Omega'\rangle\right) \tag{5}$$

where we have used the fact that $\left|\langle\Omega'|\hat{\vec{P}}|\Omega'\rangle\right|^2 \geq \langle\Omega'|\hat{\vec{P}}\cdot\vec{e}_1|\Omega'\rangle^2$ to obtain the second inequality. Obviously, then, $G\left(|\Omega'\rangle\right)$ is negative if $\langle\Omega'|\hat{F}|\Omega'\rangle$ is negative. Therefore if the energy-momentum is time-like then the condition given in (3) must be valid. It will be shown in the following discussion that quantum states must exist where the condition given in (3) does not hold which means that there are states for which the energy-momentum is space-like.

## 2. Energy-Momentum of Dirac-Maxwell system

Let the operator for the energy-momentum tensor be represented by $\hat{\theta}^{\mu\nu}(\vec{x})$. The Hamiltonian operator is then given by,

$$\hat{H} = \int \hat{\theta}^{00}(\vec{x})d\vec{x} \tag{6}$$

where $\hat{\theta}^{00}(\vec{x})$ is the operator for the energy density. The momentum operator is given by,

$$\hat{\vec{P}} = \int \left( \vec{e}_1 \hat{\theta}^{10}(\vec{x}) + \vec{e}_2 \hat{\theta}^{20}(\vec{x}) + \vec{e}_3 \hat{\theta}^{30}(\vec{x}) \right) d\vec{x} \tag{7}$$

where $\hat{\theta}^{i0}(\vec{x})$ is the operator for the momentum density in the i-th direction.

It is shown in Appendix 1 that the Hamiltonian operator $\hat{H}$ of the combined Dirac-Maxwell system in the coulomb gauge is given by

$$\hat{H} = \hat{H}_{0,D} + \hat{H}_I + \hat{H}_{CT} + \hat{H}_{0,EM} + \varepsilon_R \tag{8}$$

where,

$$\hat{H}_{0,D} = \int \left\{ -i\hat{\psi}^\dagger \vec{\alpha} \cdot \vec{\nabla}\hat{\psi} + m\hat{\psi}^\dagger \beta \hat{\psi} \right\} d\vec{x} \tag{9}$$

$$\hat{H}_I = -\int \hat{\vec{J}} \cdot \hat{\vec{A}} d\vec{x} \tag{10}$$

$$\hat{H}_{CT} = \frac{1}{2} \int d\vec{x} \int d\vec{x}' \frac{\hat{\rho}(\vec{x}) \hat{\rho}(\vec{x}')}{4\pi |\vec{x} - \vec{x}'|} \tag{11}$$

$$\hat{H}_{0,EM} = \frac{1}{2} \int \left( \hat{\vec{E}}_\perp^2 + \hat{\vec{B}}^2 \right) d\vec{x} \tag{12}$$

and where

$$\hat{\vec{B}} = \vec{\nabla} \times \hat{\vec{A}}; \quad \hat{\vec{J}} = \hat{\psi}^\dagger \vec{\alpha} \hat{\psi}; \quad \hat{\rho} = \hat{\psi}^\dagger \hat{\psi} \tag{13}$$

In the above $\hat{\psi}$ is the Dirac field operator, $\vec{\alpha}$ and $\beta$ are the usual 4x4 matrices, $\hat{\vec{J}}$ is the current operator, $\hat{\rho}$ is the charge operator, $\hat{\vec{A}}$ is the operator for the vector potential, $\hat{\vec{E}}_\perp$ is the transverse electric field operator, $\hat{\vec{B}}$ is the magnetic field operator, and $\varepsilon_R$ is a renormalization constant so that the energy of the vacuum state is zero. Natural units are used so that $\hbar = c = 1$. Also the Schrödinger picture is assumed so that field operators are dependent on space. The time dependence is carried by the state vector.

The momentum operator of the Dirac-Maxwell field is shown in Appendix 1 to be,

$$\hat{\vec{P}} = \hat{\vec{P}}_{0,D} + \hat{\vec{P}}_{0,EM} \qquad (14)$$

where,

$$\hat{\vec{P}}_{0,D} = -i\int \psi^{\dagger}\vec{\nabla}\psi\, d\vec{x} \qquad (15)$$

$$\hat{\vec{P}}_{0,EM} = \int \left(\hat{\vec{E}}_{\perp} \times \hat{\vec{B}}\right) d\vec{x} \qquad (16)$$

For the coulomb gauge the commutator relationships are given by Ryder [4] as

$$\left[\hat{A}^{i}(\vec{x}), \hat{E}_{\perp}^{j}(\vec{x}')\right] = i\int \frac{d\vec{k}}{(2\pi)^{3}}\left(\delta^{ij} - \frac{k^{i}k^{j}}{|\vec{k}|^{2}}\right)e^{i\vec{k}\cdot(\vec{x}-\vec{x}')} \qquad (17)$$

and

$$\left[\hat{E}_{\perp}^{i}(\vec{x}'), \hat{E}_{\perp}^{j}(\vec{x})\right] = \left[\hat{A}^{i}(\vec{x}'), \hat{A}^{j}(\vec{x})\right] = 0 \qquad (18)$$

In addition, the Maxwell field operators $\hat{\vec{E}}_{\perp}(\vec{x})$ and $\hat{\vec{A}}(\vec{x})$ commute with the Dirac field operators $\hat{\psi}^{\dagger}$ and $\hat{\psi}$.

Now given a normalized state vector $|\Omega\rangle$ we can always define another state vector by acting on $|\Omega\rangle$ with a function of the field operators [3]. Therefore we can define the state vector $|\Omega'\rangle$ by the expression

$$|\Omega'\rangle = e^{i\hat{C}_{1}}e^{i\hat{C}_{2}}|\Omega\rangle \qquad (19)$$

where operators $\hat{C}_{1}$ and $\hat{C}_{2}$ are defined by

$$\hat{C}_1 = -\int \hat{\vec{E}}_\perp \cdot \vec{A}_{cl} d\vec{x} \tag{20}$$

and

$$\hat{C}_2 = \int \hat{\vec{A}} \cdot \vec{E}_{cl} d\vec{x} \tag{21}$$

and where $\vec{A}_{cl}(\vec{x})$ and $\vec{E}_{cl}(\vec{x})$ are real-valued vector functions whose divergence is zero, i.e.,

$$\vec{\nabla} \cdot \vec{A}_{cl}(\vec{x}) = \vec{\nabla} \cdot \vec{E}_{cl}(\vec{x}) = 0 \tag{22}$$

Note that $\hat{\vec{E}}_\perp$ and $\hat{\vec{A}}$ are real operators so that $\hat{\vec{A}} = \hat{\vec{A}}^\dagger$ and $\hat{\vec{E}}_\perp = \hat{\vec{E}}_\perp^\dagger$. This yields $\hat{C}_1^\dagger = \hat{C}_1$ and $\hat{C}_2^\dagger = \hat{C}_2$ so that,

$$e^{-i\hat{C}_1^\dagger} e^{i\hat{C}_1} = e^{-i\hat{C}_2^\dagger} e^{i\hat{C}_2} = 1 \text{ and } \langle \Omega'|\Omega'\rangle = \langle \Omega|\Omega\rangle \tag{23}$$

From the above expressions it shown in Appendix 2 that

$$\left[\hat{\vec{A}}(\vec{x}), \hat{C}_1\right] = -i\vec{A}_{cl}(\vec{x}) \text{ and } \left[\hat{\vec{E}}_\perp(\vec{x}), \hat{C}_2\right] = -i\vec{E}_{cl}(\vec{x}) \tag{24}$$

Next we will use the Baker-Campell-Hausdorff relationships [5] which state that,

$$e^{+\hat{O}_1} \hat{O}_2 e^{-\hat{O}_1} = \hat{O}_2 + \left[\hat{O}_1, \hat{O}_2\right] + \frac{1}{2}\left[\hat{O}_1, \left[\hat{O}_1, \hat{O}_2\right]\right] + \ldots \tag{25}$$

where $\hat{O}_1$ and $\hat{O}_2$ are operators. Use these relationships and the fact that the fact that $\left[\hat{\vec{A}}(\vec{x}), \hat{C}_1\right]$ is a c-number so that it commutes with $\hat{C}_1$ to obtain,

$$e^{-i\hat{C}_1} \hat{\vec{A}}(\vec{x}) e^{i\hat{C}_1} = \hat{\vec{A}}(\vec{x}) + i\left[\hat{\vec{A}}(\vec{x}), \hat{C}_1\right] = \hat{\vec{A}}(\vec{x}) + \vec{A}_{cl}(\vec{x}) \tag{26}$$

Similarly

$$e^{-i\hat{C}_2^\dagger} \hat{\vec{E}}_\perp e^{i\hat{C}_2} = \hat{\vec{E}}_\perp + \vec{E}_{cl}(\vec{x}) \tag{27}$$

Also

$$e^{-i\hat{C}_1^{\dagger}}\hat{\vec{A}}(\vec{x})\hat{\vec{A}}(\vec{x}')e^{i\hat{C}_1} = e^{-i\hat{C}_1^{\dagger}}\hat{\vec{A}}(\vec{x})e^{i\hat{C}_1}e^{-i\hat{C}_1^{\dagger}}\hat{\vec{A}}(\vec{x}')e^{i\hat{C}_1}$$
$$= \left(\hat{\vec{A}}(\vec{x}) + \vec{A}_{cl}(\vec{x})\right)\left(\hat{\vec{A}}(\vec{x}') + \vec{A}_{cl}(\vec{x}')\right) \quad (28)$$

Similarly

$$e^{-i\hat{C}_2^{\dagger}}\hat{\vec{E}}_{\perp}(\vec{x})\hat{\vec{E}}_{\perp}(\vec{x}')e^{i\hat{C}_2} = \left(\hat{\vec{E}}_{\perp}(\vec{x}) + \vec{E}_{cl}(\vec{x})\right)\left(\hat{\vec{E}}_{\perp}(\vec{x}') + \vec{E}_{cl}(\vec{x}')\right) \quad (29)$$

Let $\hat{O}_D = \hat{\vec{J}}, \hat{H}_{0,D}, \hat{\rho}$, or $\hat{P}_{0,D}$. Since these quantities are functions of Dirac field operators they commute with $\hat{C}_1$ and $\hat{C}_2$. Therefore we have,

$$e^{-i\hat{C}_1^{\dagger}}\hat{O}_D e^{i\hat{C}_1} = \hat{O}_D \text{ and } e^{-i\hat{C}_2^{\dagger}}\hat{O}_D e^{i\hat{C}_2} = \hat{O}_D \quad (30)$$

### **3. Space-like energy-momentum**

In the previous section we developed a number of expressions that will be useful in evaluating the momentum and energy of the state $|\Omega'\rangle$ as defined in (19). First determine the momentum $\langle\Omega'|\hat{\vec{P}}(\hat{\vec{A}},\hat{\vec{E}}_{\perp})|\Omega'\rangle$ where we write $\hat{\vec{P}}(\hat{\vec{A}},\hat{\vec{E}}_{\perp})$ to explicitly show the dependence of the operator $\hat{\vec{P}}$ on the operators $\hat{\vec{A}}$ and $\hat{\vec{E}}_{\perp}$.

From (19), (14) through (16), and (26) through (30) we obtain

$$\langle\Omega'|\hat{\vec{P}}(\hat{\vec{A}},\hat{\vec{E}}_{\perp})|\Omega'\rangle = \langle\Omega|e^{-i\hat{C}_2^{\dagger}}e^{-i\hat{C}_1^{\dagger}}\hat{\vec{P}}(\hat{\vec{A}},\hat{\vec{E}}_{\perp})e^{i\hat{C}_1}e^{i\hat{C}_2}|\Omega\rangle$$
$$= \langle\Omega|\hat{\vec{P}}(\hat{\vec{A}}+\vec{A}_{cl},\hat{\vec{E}}_{\perp}+\vec{E}_{cl})|\Omega\rangle \quad (31)$$

Use (31) along with (14) through (16) to obtain,

$$\langle\Omega'|\hat{\vec{P}}|\Omega'\rangle = \langle\Omega|\hat{\vec{P}}_{0,D}|\Omega\rangle + \langle\Omega|\int\left[\left(\hat{\vec{E}}_{\perp}+\vec{E}_{cl}\right)\times\left(\hat{\vec{B}}+\vec{B}_{cl}\right)\right]d\vec{x}|\Omega\rangle \quad (32)$$

where

$$\vec{B}_{cl} = \vec{\nabla} \times \vec{A}_{cl} \tag{33}$$

This yields

$$\langle \Omega' | \hat{\vec{P}} | \Omega' \rangle = \langle \Omega | \hat{\vec{P}} | \Omega \rangle + \int \left\{ \left( \vec{E}_e \times \vec{B}_{cl} \right) + \left( \vec{E}_{cl} \times \vec{B}_e \right) \right\} d\vec{x} + \int \left( \vec{E}_{cl} \times \vec{B}_{cl} \right) d\vec{x} \tag{34}$$

where we have defined the expectation values,

$$\vec{E}_e = \langle \Omega | \hat{\vec{E}}_\perp | \Omega \rangle \text{ and } \vec{B}_e = \langle \Omega | \hat{\vec{B}} | \Omega \rangle \tag{35}$$

Next calculate the energy of the state $|\Omega'\rangle$. Based on the previous discussion this is given by

$$\langle \Omega' | \hat{H}\left(\hat{\vec{A}}, \hat{\vec{E}}_\perp\right) | \Omega' \rangle = \langle \Omega | \hat{H}\left(\hat{\vec{A}} + \vec{A}_{cl}, \hat{\vec{E}}_\perp + \vec{E}_{cl}\right) | \Omega \rangle \tag{36}$$

Therefore, we obtain

$$\langle \Omega' | \hat{H} | \Omega' \rangle = \langle \Omega | \hat{H} | \Omega \rangle + \int \left\{ \vec{E}_e \cdot \vec{E}_{cl} + \vec{B}_e \cdot \vec{B}_{cl} \right\} d\vec{x} \\ - \int \vec{J}_e \cdot \vec{A}_{cl} d\vec{x} + \frac{1}{2} \int \left( \vec{E}_{cl}^{\,2} + \vec{B}_{cl}^{\,2} \right) d\vec{x} \tag{37}$$

where the current expectation value $\vec{J}_e$ is defined by

$$\vec{J}_e = \langle \Omega | \hat{\vec{J}} | \Omega \rangle \tag{38}$$

Eq. (37) can be written as,

$$\langle \Omega' | \hat{H} | \Omega' \rangle = \langle \Omega | \hat{H} | \Omega \rangle + \eta + \xi_{cl} - \int \vec{J}_e \cdot \vec{A}_{cl} d\vec{x} \tag{39}$$

where,

$$\eta = \int \left\{ \vec{E}_e \cdot \vec{E}_{cl} + \vec{B}_e \cdot \vec{B}_{cl} \right\} d\vec{x} \tag{40}$$

and

$$\xi_{cl} = \frac{1}{2} \int \left( \vec{E}_{cl}^{\,2} + \vec{B}_{cl}^{\,2} \right) d\vec{x} \tag{41}$$

Similarly (34) can be expressed as,

$$\langle\Omega'|\hat{\vec{P}}|\Omega'\rangle = \langle\Omega|\hat{\vec{P}}|\Omega\rangle + \vec{\omega} + \vec{P}_{cl} \tag{42}$$

where,

$$\vec{\omega} = \int\left\{\left(\vec{E}_e \times \vec{B}_{cl}\right) + \left(\vec{E}_{cl} \times \vec{B}_e\right)\right\}d\vec{x} \tag{43}$$

and

$$\vec{P}_{cl} = \int\left(\vec{E}_{cl} \times \vec{B}_{cl}\right)d\vec{x} \tag{44}$$

Use (42), (39), and (2) to obtain

$$\langle\Omega'|\hat{F}|\Omega'\rangle = \langle\Omega|\hat{F}|\Omega\rangle + \left(\eta - \vec{\omega}\cdot\vec{e}_1\right) + F_{cl} - \int\vec{J}_e \cdot \vec{A}_{cl}d\vec{x} \tag{45}$$

where,

$$F_{cl} = \xi_{cl} - \vec{P}_{cl}\cdot\vec{e}_1 \tag{46}$$

At this point we want to try to find an $\vec{A}_{cl}$ and $\vec{E}_{cl}$ which makes $\langle\Omega'|\hat{F}|\Omega'\rangle$ negative. One possible solution is,

$$\vec{A}_{cl} = \vec{e}_2 fg(x_1); \quad \vec{E}_{cl} = \vec{e}_2 f\frac{\partial g(x_1)}{\partial x_1} \tag{47}$$

where 'f' is a positive constant and $g(x_1)$ is an arbitrary function that is dependent only on the $x_1$ coordinate. This solution obviously satisfies (22).

From this we obtain,

$$\vec{B}_{cl} = \nabla \times \vec{A}_{cl} = \vec{e}_3 f\frac{\partial g(x_1)}{\partial x_1} \tag{48}$$

Let $B_{cl}^{(i)}$ be the i-th component of $\vec{B}_{cl}$. From the above it is evident that

$$B_{cl}^{(3)} = f\,\partial g(x_1)/\partial x_1 \text{ and that } E_{cl}^{(2)} = B_{cl}^{(3)}. \tag{49}$$

Use these results to obtain,

$$\vec{e}_1 \cdot (\vec{E}_{cl} \times \vec{B}_{cl}) = (B_{cl}^{(3)})^2 \text{ and } \frac{1}{2}(\vec{E}_{cl}^2 + \vec{B}_{cl}^2) = (B_{cl}^{(3)})^2 \tag{50}$$

Use this in (46) to obtain $F_{cl} = 0$. We can also show that.

$$\vec{E}_e \cdot \vec{E}_{cl} + \vec{B}_e \cdot \vec{B}_{cl} = (E_e^{(2)} + B_e^{(3)}) B_{cl}^{(3)} \tag{51}$$

and,

$$\vec{e}_1 \cdot \{(\vec{E}_e \times \vec{B}_{cl}) + (\vec{E}_{cl} \times \vec{B}_e)\} = (E_e^{(2)} + B_e^{(3)}) B_{cl}^{(3)} \tag{52}$$

From this we obtain $(\eta - \vec{\omega} \cdot \vec{e}_1) = 0$

Use these results in (45) to obtain,

$$\langle \Omega' | \hat{F} | \Omega' \rangle = \langle \Omega | \hat{F} | \Omega \rangle - f \int \vec{J}_e \cdot \vec{e}_2 g(x_1) d\vec{x} \tag{53}$$

Now assume that we have chosen the state $|\Omega\rangle$ so that

$$\int \vec{J}_e \cdot \vec{e}_2 g(x_1) d\vec{x} \neq 0 \tag{54}$$

Now how do we know that a state $|\Omega\rangle$ exists where this is true? Recall that $g(x_1)$ is an arbitrary function. If our theory is a correct model of the real world then states must exist where this condition holds for some $g(x_1)$ because the above relationship obviously can be true in classical physics which is often a very close approximation to the real world.

The constant f does not appear in the expression $\langle \Omega | \hat{F} | \Omega \rangle$. Therefore, given (54), we can always find an f where

$$\langle \Omega' | \hat{F} | \Omega' \rangle < 0 \tag{55}$$

which means that the energy-momentum of the state $|\Omega'\rangle$ is space-like.

## 4. The boundary conditions

Now there is a possible problem with the electric potential of (47) involving the boundary conditions at infinity. We can always specify the function $g(x_1)$ so that $\vec{A}_{cl}$ and $\vec{E}_{cl}$ go to zero as $x_1 \to \pm\infty$. However $\vec{A}_{cl}$ and $\vec{E}_{cl}$ are dependent only on the coordinate $x_1$ so they will not go to zero as $x_2$ or $x_3 \to \pm\infty$. In order to deal with this objection we will rework the problem for the case where $\vec{A}_{cl}$ and $\vec{E}_{cl}$ go to zero at infinity in all directions.

To do this specify $\vec{A}_{cl}$ as follows,

$$\vec{A}_{cl} = \text{fr}(x_2, x_3)\left(ax_2 \frac{\partial h(x_1)}{\partial x_1}, g(x_1), 0\right) \tag{56}$$

where,

$$r(x_2, x_3) = \exp\left(-\frac{a}{2}\left(x_2^2 + x_3^2\right)\right) \tag{57}$$

and 'a' is a positive constant, 'f' is a constant, and $g(x_1) = \partial^2 h(x_1)/\partial x_1^2$ where $h(x_1) \to 0$ as $x_1 \to \pm\infty$. Other than this $h(x_1)$ is an arbitrary function of the $x_1$ coordinate.

From the above we obtain,

$$\vec{B}_{cl} = \nabla \times \vec{A}_{cl} = \left(-\frac{\partial A_{cl}^{(2)}}{\partial x_3}, \frac{\partial A_{cl}^{(1)}}{\partial x_3}, \left(\frac{\partial A_{cl}^{(2)}}{\partial x_1} - \frac{\partial A_{cl}^{(1)}}{\partial x_2}\right)\right) \tag{58}$$

Next specify $\vec{E}_{cl}$ as follows,

$$\vec{E}_{cl} = \left(E_{cl}^{(1)}, B_{cl}^{(3)}, 0\right) \tag{59}$$

where $B_{cl}^{(3)}$ is obtained by referring to (58) and $E_{cl}^{(1)}$ is given by,

$$E_{cl}^{(1)} = -\frac{\partial A_{cl}^{(2)}}{\partial x_2} + fah\frac{\partial^2 (x_2 r)}{\partial x_2^2} \tag{60}$$

Note that $\vec{A}_{cl}$ and $\vec{E}_{cl}$ satisfy (22) and go to zero at infinity. Use the above in (46) to show that,

$$F_{cl} = \frac{1}{2}\int\left\{\left(\frac{\partial A_{cl}^{(2)}}{\partial x_3}\right)^2 + \left(\frac{\partial A_{cl}^{(1)}}{\partial x_3}\right)^2 + \left(E_{cl}^{(1)}\right)^2\right\}d\vec{x} \tag{61}$$

Also using the above in (40) and (43) to obtain,

$$\eta - \vec{\omega}\cdot\vec{e}_1 = \int\left\{E_e^{(1)}E_{cl}^{(1)} - B_e^{(1)}\left(\frac{\partial A_{cl}^{(2)}}{\partial x_3}\right) + \left(B_e^{(2)} + E_e^{(3)}\right)\left(\frac{\partial A_{cl}^{(1)}}{\partial x_3}\right)\right\}d\vec{x} \tag{62}$$

In order to evaluate these relationships use the following,

$$\frac{\partial A_{cl}^{(2)}}{\partial x_3} = -ax_3 A_{cl}^{(2)} \tag{63}$$

$$\frac{\partial A_{cl}^{(1)}}{\partial x_3} = -ax_3 A_{cl}^{(1)} \tag{64}$$

$$E_{cl}^{(1)} = afr\left(x_2 g + h\left(a^2 x_2^3 - 3ax_2\right)\right) \tag{65}$$

Now we will evaluate (61) in the limit $a \to 0$. In this case it can be shown that,

$$F_{cl}\underset{a\to 0}{=}\left(\frac{\pi f^2}{4}\right)R \tag{66}$$

where,

$$R = \int_{-\infty}^{+\infty}\left(\frac{\partial^2 h}{\partial x_1^2}\right)^2 dx_1 \tag{67}$$

Since h is arbitrary we assume we specify an h where R is finite.

Now let us examine the situation so far. Consider the volume of space $V_L(a)$ specified by $|x_2| \leq L$ and $|x_3| \leq L$ where $L = 1/\left(a^{(1/4)}\right)$. We can see that in this region the Gaussian function $r(x_2, x_3) \simeq 1$ if 'a' is small. Therefore, as $a \to 0$, the expressions for $\vec{A}_{cl}$, $\vec{B}_{cl}$, and $\vec{E}_{cl}$ in the region $V_L(a)$ are essentially the same as in Section 3. However there is important difference between these results and those of the previous section which is that in this case $F_{cl}$ is not zero. This can be thought of as an effect of imposing the boundary condition that the various quantities must go to zero at infinity. This complicates the discussion somewhat. In order to proceed we will use the fact that as 'a' becomes smaller the volume $V_L(a)$ increases while $F_{cl}$ remains constant. Therefore we want to consider an example where the increasing volume effect overwhelms the constant boundary effect.

In order to do this we will make reasonable assumptions concerning the expectation values $\theta_e^{\mu\nu} = \langle \Omega | \hat{\theta}^{\mu\nu} | \Omega \rangle$, $\vec{J}_e$, $\vec{B}_e$, and $\vec{E}_e$ of the quantum state $|\Omega\rangle$. The main assumption is a reasonable correspondence between quantum mechanics and classical physics in situations where classical physics applies. That is, although we believe that quantum mechanics provides the correct mathematical description of the physical world there are many situations where the classical physics can describe a system to a high degree of accuracy. Due to the fact that the actual correct mathematical description is based on quantum mechanics there must be a correspondence between the quantum mechanical state vector and the observables of the classical system. This is done at the level of expectation values. For example if the magnetic field in the classical system has the value $\vec{B}(\vec{x})$ and if $|\Omega\rangle$ is the quantum state that gives the quantum mechanical description of this system then $\vec{B}_e(\vec{x}) = \langle\Omega|\hat{\vec{B}}(\vec{x})|\Omega\rangle \simeq \vec{B}(\vec{x})$. Similar relationships hold for the other observables.

Now if we examine (45) we can see that all we care about at this point are the expectation values associated with the state vector $|\Omega\rangle$. The actual form of the state vector is not needed. Therefore statements that we can make about the observables of a classical system can be expected to apply to the expectation values of the quantum system.

Now consider a classical system that is confined to a finite volume $V_1$. That is all the parts of the system including the electromagnetic field, current and charge density, and the energy density and momentum density are non-zero within the volume $V_1$ and are zero or approach zero very rapidly outside $V_1$. Now consider another volume $V_2$ that is separated from $V_1$ by a very large distance. Let another classical system be confined volume $V_2$. In fact we can assume that the system within $V_2$ is identical to that within $V_1$. In this case the total energy and momentum of the total system is just twice the energy and momentum of each separate system. We can of course add additional systems in separate and isolated volumes of space. The total energy-momentum is the sum of the energy-momentum of the separate systems. Let us designate these volumes by $V_n$ where $n = 1$ to $N$. Further more we assume that all the $V_n$ are within the volume $V_L(a)$.

To sum up we consider a classical system consisting of N identical systems each confined to a volume $V_n$. Each of the subsystems has a energy-momentum, current, charge, and electromagnetic field that is confined to $V_n$. Now consider the quantum state $|\Omega\rangle$ that corresponds to this classical system. Let $O_e(\vec{x})$ refer to any one of the expectation values $\theta_e^{\mu\nu}$, $\vec{J}_e$, $\rho_e$, $\vec{B}_e$, and $\vec{E}_e$ of the quantum state $|\Omega\rangle$. The $O_e(\vec{x})$ must be very close in value to the observables of the classical system. Therefore we assume that the $O_e(\vec{x})$ take on non-zero values only in the sufficiently separate volumes $V_n$ where $n = 1$ to $N$. In this case (45) becomes,

$$\langle \Omega' | \hat{F} | \Omega' \rangle = F_{cl} + \sum_{n=1}^{N} S_n \tag{68}$$

where,

$$S_n \equiv \left\{ F_e(V_n) + \left( \eta(V_n) - \vec{\omega}(V_n) \cdot \vec{e}_1 \right) - \int_{V_n} \vec{J}_e \cdot \vec{A}_{cl} d\vec{x} \right\} \tag{69}$$

and

$$F_e(V_n) = \int_{V_n} \left( \theta_e^{00}(\vec{x}) - \theta_e^{10}(\vec{x}) \right) d\vec{x} \tag{70}$$

$$\eta(V_n) = \int_{V_n} \left\{ \vec{E}_e \cdot \vec{E}_{cl} + \vec{B}_e \cdot \vec{B}_{cl} \right\} d\vec{x} \tag{71}$$

$$\vec{\omega}(V_n) = \int_{V_n} \left\{ \left( \vec{E}_e \times \vec{B}_{cl} \right) + \left( \vec{E}_{cl} \times \vec{B}_e \right) \right\} d\vec{x} \tag{72}$$

The subscript $V_n$ under the integral sign means that the integration is over the volume $V_n$. Now we assume that all the $V_n$ are within the region $V_L(a)$. This is always possible because as $a \to 0$ $V_L(a)$ becomes arbitrarily large. This allows us to easily evaluate $\eta(V_n) - \vec{\omega}(V_n) \cdot \vec{e}_1$. By examining (62) through (65) it is evident that the integrand in (62) is proportional 'a' in the limit that $a \to 0$. Since the volume $V_n$ is finite $\eta(V_n) - \vec{\omega}(V_n) \cdot \vec{e}_1$ is proportional 'a' and therefore will approach zero as $a \to 0$. Therefore the term $\eta(V_n) - \vec{\omega}(V_n) \cdot \vec{e}_1$ can be eliminated from the expression for $S_n$ to obtain,

$$S_n = \left\{ F_e(V_n) - \int_{V_n} \vec{J}_e \cdot \vec{A}_{cl} d\vec{x} \right\} \tag{73}$$

We can rewrite this as,

$$S_n = \left\{ F_e(V_n) - f \int_{V_n} \vec{J}_e \cdot \vec{A}'_{cl} d\vec{x} \right\} \tag{74}$$

where $\vec{A}'_{cl} = \vec{A}_{cl}/f$. Referring to (56) we can see that $\vec{A}'_{cl}$ is independent of f. Also $F_e(V_n)$ is independent of f. Therefore if $\int_{V_n} \vec{J}_e \cdot \vec{A}'_{cl} d\vec{x}$ is non-zero we can always make f large enough so that $S_n < 0$. Refer to (68) and use the fact that the $S_n$ are all identical and negative to obtain,

$$\langle \Omega' | \hat{F} | \Omega' \rangle = F_{cl} - N |S_n| \tag{75}$$

Use (66) in the above to obtain,

$$\langle \Omega' | \hat{F} | \Omega' \rangle \underset{a \to 0}{=} \left( \frac{\pi f^2}{4} \right) R - N |S_n| \tag{76}$$

At this point it is not clear if the above quantity can be made negative. To show that it can be we will examine the dependence of the terms on the right hand side of the equation on 'a' as $a \to 0$. The first term is obviously independent of 'a' as is $S_n$. Now consider N. N is the number of sub volumes $V_n$ that can fit in $V_L(a)$. It is obviously proportional to the size of $V_L(a)$. From the definition of $V_L(a)$ this yields $N \sim 1/\sqrt{a}$. Therefore as $a \to 0$ N can be made arbitrarily large so that $\langle \Omega' | \hat{F} | \Omega' \rangle$ will be negative.

In conclusion, we have examined the commonly held assumption that the energy-momentum 4-vector of any quantum state must be time-like. We have considered the case of the Dirac-Maxwell field and have shown that there must exist quantum states for which the energy-momentum must be space-like.

## Appendix 1

The energy momentum tensor $\theta^{\mu\nu}$ of the Dirac-Maxwell field is discussed in Greiner [5] (see equation 5 on page 151 of this reference). From this we can derive the total energy and momentum as

$$\hat{P}^0 = \hat{H}_{0,D} - \int \hat{\vec{J}} \cdot \hat{\vec{A}} d\vec{x} + \frac{1}{2} \int \left( \hat{\vec{E}}^2 + \hat{\vec{B}}^2 \right) d\vec{x} \tag{A1-1}$$

and

$$\hat{\vec{P}} = \hat{\vec{P}}_{0,D} - \int \hat{\rho} \hat{\vec{A}} d\vec{x} + \int \left( \hat{\vec{E}} \times \hat{\vec{B}} \right) d\vec{x} \tag{A1-2}$$

respectively (See chapters 5 and 6 of [6] and Appendix A of Sakurai [6]). The electromagnetic field will be quantized in the coulomb gauge. In this case

$$\vec{\nabla} \cdot \hat{\vec{A}} = 0 \tag{A1-3}$$

where

$$\hat{\vec{E}} = \left( \hat{\vec{E}}_\perp - \vec{\nabla} \hat{A}_0 \right) \text{ and } \hat{\vec{B}} = \vec{\nabla} \times \hat{\vec{A}} \tag{A1-4}$$

$\hat{\vec{E}}_\perp$ is the transverse component of the electric field and corresponds to the $-\partial \vec{A}/\partial t$ term of the classical unquantized electric field. $\hat{\vec{E}}_\perp$ satisfies

$$\vec{\nabla} \cdot \hat{\vec{E}}_\perp = 0 \tag{A1-5}$$

In the coulomb gauge it can be shown that

$$\vec{\nabla}^2 \hat{A}_0 = -\hat{\rho} \tag{A1-6}$$

From Appendix A of Sakurai [6] it is shown that,

$$\frac{1}{2}\int\left(\hat{\vec{E}}^2 + \hat{\vec{B}}^2\right)d\vec{x} = \frac{1}{2}\int\left(\hat{\vec{E}}_\perp^2 + \hat{\vec{B}}^2\right)d\vec{x} + \frac{1}{2}\int d\vec{x}\int d\vec{x}' \frac{\hat{\rho}(\vec{x})\hat{\rho}(\vec{x}')}{4\pi|\vec{x}-\vec{x}'|} \tag{A1-7}$$

Use this in (A1-1). Then replace $\hat{P}^0$ by $\hat{H}$ to convert to the notation used in this discussion and add the renormalization constant $\varepsilon_R$ to obtain (8).

Next, use (A1-4) to obtain,

$$\int\left(\hat{\vec{E}} \times \hat{\vec{B}}\right)d\vec{x} = \int\left(\left(\hat{\vec{E}}_\perp - \vec{\nabla}\hat{A}_0\right) \times \hat{\vec{B}}\right)d\vec{x} \tag{A1-8}$$

Next evaluate $\int\left(\vec{\nabla}\hat{A}_0 \times \hat{\vec{B}}\right)d\vec{x} = \int\left(\vec{\nabla}\hat{A}_0 \times \left(\vec{\nabla} \times \hat{\vec{A}}\right)\right)d\vec{x}$. From a vector identity,

$$\vec{\nabla}\left(\hat{\vec{A}} \cdot \vec{\nabla}\hat{A}_0\right) = \left(\hat{\vec{A}} \cdot \vec{\nabla}\right)\vec{\nabla}\hat{A}_0 + \left(\vec{\nabla}\hat{A}_0 \cdot \vec{\nabla}\right)\hat{\vec{A}} + \vec{\nabla}\hat{A}_0 \times \left(\vec{\nabla} \times \hat{\vec{A}}\right) \tag{A1-9}$$

we obtain

$$\int\vec{\nabla}\hat{A}_0 \times \left(\vec{\nabla} \times \hat{\vec{A}}\right)d\vec{x} = -\int\left\{\left(\vec{\nabla}\hat{A}_0 \cdot \vec{\nabla}\right)\hat{\vec{A}} + \left(\hat{\vec{A}} \cdot \vec{\nabla}\right)\vec{\nabla}\hat{A}_0\right\}d\vec{x} \tag{A1-10}$$

where we have assumed reasonable boundary conditions so that

$$\int\vec{\nabla}\left(\hat{\vec{A}} \cdot \vec{\nabla}\hat{A}_0\right)d\vec{x} = 0 \tag{A1-11}$$

Assume reasonable boundary conditions and integrate by parts to obtain

$$\int \vec{\nabla}\hat{A}_0 \times \left(\vec{\nabla} \times \hat{\vec{A}}\right) d\vec{x} = \int \left\{ \left(\vec{\nabla}^2 \hat{A}_0\right) \hat{\vec{A}} + \left(\vec{\nabla} \cdot \hat{\vec{A}}\right) \vec{\nabla}\hat{A}_0 \right\} d\vec{x} \tag{A1-12}$$

Use (A1-3) and (A1-6) in the above expression to obtain

$$\int \vec{\nabla}\hat{A}_0 \times \left(\vec{\nabla} \times \hat{\vec{A}}\right) d\vec{x} = -\int \hat{\rho}\hat{\vec{A}} d\vec{x} \tag{A1-13}$$

Use the above expressions to obtain

$$\int \left(\hat{\vec{E}} \times \hat{\vec{B}}\right) d\vec{x} = \int \left(\hat{\vec{E}}_\perp \times \hat{\vec{B}}\right) d\vec{x} + \int \hat{\rho}\hat{\vec{A}} d\vec{x} \tag{A1-14}$$

Use this in (A1-2) to obtain (14) in the text.

## Appendix 2

From the definition of $\hat{C}_1$

$$\left[\hat{\vec{A}}(\vec{x}), \hat{C}_1\right] = -\int \left[\hat{\vec{A}}(\vec{x}), \hat{\vec{E}}_\perp(\vec{x}')\right] \cdot \vec{A}_{cl}(\vec{x}') d\vec{x}' \tag{A2-1}$$

Use (17) in the above to obtain

$$\left[\hat{\vec{A}}(\vec{x}), \hat{C}_1\right] = -i\int d\vec{x}' \int \frac{d\vec{k}}{(2\pi)^3} \left( \vec{A}_{cl}(\vec{x}') - \frac{\vec{k}\left(\vec{k} \cdot \vec{A}_{cl}(\vec{x}')\right)}{\left|\vec{k}\right|^2} \right) e^{i\vec{k}\cdot(\vec{x}-\vec{x}')} \tag{A2-2}$$

Use,

$$\int \frac{d\vec{k}}{(2\pi)^3} e^{i\vec{k}\cdot(\vec{x}-\vec{x}')} = \delta^{(3)}(\vec{x} - \vec{x}')$$

to obtain

$$\left[\hat{\vec{A}}(\vec{x}), \hat{C}_1\right] = -i\vec{A}_{cl}(\vec{x}) + i\int d\vec{x}' \int \frac{d\vec{k}}{(2\pi)^3} \frac{\vec{k}\left(\vec{k} \cdot \vec{A}_{cl}(\vec{x}')\right)}{\left|\vec{k}\right|^2} e^{i\vec{k}\cdot(\vec{x}-\vec{x}')} \tag{A2-3}$$

It can be easily shown that if $\vec{\nabla} \cdot \vec{A}_{cl}(\vec{x}') = 0$ that the second term to the left of the equals sign is zero. Therefore $\left[ \hat{\vec{A}}(\vec{x}), \hat{C}_1 \right] = -i\vec{A}_{cl}(\vec{x})$. Similarly it can be shown that

$$\left[ \hat{\vec{E}}_\perp(\vec{x}), \hat{C}_2 \right] = -i\vec{E}_{cl}(\vec{x}).$$

## References

1. R. Haag, *Local Quantum Physics*, Springer-Velag, Berlin (1993).

2. K. Nishijima, *Field and Particles: Field theory and Dispersion Relationships,* W.A. Benjamin, New York, (1969).

3. S. Weinberg, *The Quantum Theory of Fields Vol. 1*, Press Syndicate of the University of Cambridge, Cambridge, (1995).

4. L. H. Ryder, *Quantum Field Theory*, Cambridge University Press, Cambridge, (1985).

5. W. Greiner and J. Reinhardt. *Quantum Electrodynamics*. Springer-Verlag, Berlin. (1992).

6. J. J. Sakurai, *Advanced Quantum Mechanics* , Addison-Wesley Publishing Co., Redwood, Calif (1967).